\documentclass{appolb}
\pdfoutput=1
\usepackage{graphicx,wrapfig,float,slashed,indentfirst}
\usepackage{amsmath,amssymb,epsfig,xcolor}
\usepackage{fancyhdr}
\usepackage{epstopdf}
\usepackage{graphics}
\usepackage[mathscr]{eucal}
\usepackage{feynmp}
\usepackage{xspace}
\usepackage{listings}
\usepackage{gensymb}
\usepackage{pstricks}
\usepackage{xspace}
\usepackage{braket}
\usepackage{subfigure}
\usepackage{array}
\usepackage{feynmp}
\newcommand{\SARAH}{{\tt SARAH}\xspace}
\newcommand{\PYRATE}{{\tt Pyrate}\xspace}

\newcommand{\nc}{\newcommand}
\nc{\beq}{\begin{equation}}  \nc{\eeq}{\end{equation}}
\nc{\bea}{\begin{eqnarray}}  \nc{\eea}{\end{eqnarray}}
\nc{\baa}{\begin{array}}     \nc{\eaa}{\end{array}}
\nc{\bit}{\begin{itemize}}   \nc{\eit}{\end{itemize}}
\nc{\ben}{\begin{enumerate}} \nc{\een}{\end{enumerate}}
\nc{\bce}{\begin{center}}    \nc{\ece}{\end{center}}
\nc{\bpm}{\begin{pmatrix}}   \nc{\epm}{\end{pmatrix}}
\nc{\bvt}{\begin{verbatim}}  \nc{\evt}{\end{verbatim}}
\nc{\non}{\nonumber} 
\newcolumntype{M}{>{$\vcenter\bgroup\hbox\bgroup}c<{\egroup\egroup$}}

\def\gev{\;\hbox{GeV}}

\def\diag{\hbox{\diag}}
\def\zBB{{\mathbb{Z}}}

\def\z2{\zBB_2}

\def\mone{M_{h_1}}
\def\mtwo{M_{h_2}}
\def\zp{Z^\prime}
\def\mzp{M_{\zp}}
\def\vx{v_x}
\def\gx{g_x}
\def\xdd{\sigma_{\zp N}}

\def\lsim{\mathrel{\raise.3ex\hbox{$<$\kern-.75em\lower1ex\hbox{$\sim$}}}}
\def\gsim{\mathrel{\raise.3ex\hbox{$>$\kern-.75em\lower1ex\hbox{$\sim$}}}}

\def\ot#1{%
  \mathrel{\vbox{\offinterlineskip\ialign{%
    \hfil##\hfil\cr
    $\scriptscriptstyle(\,\sim\,)$\cr
    \noalign{\kern-.1ex}
    $#1$\cr
}}}}

\begin{document}
\title{Vacuum stability from vector dark matter
\thanks{Presented by M. Duch at the XXXIX International Conference of Theoretical Physics ``Matter to the deepest 2015''}}
\author{M. Duch, B. Grzadkowski, M. McGarrie
\address{Faculty of Physics, University of Warsaw, Pasteura 5, 02-093 Warsaw, Poland}}
\maketitle
\begin{abstract}
 We study a model of vector dark matter with the complex scalar Higgs portal. Renormalisation group equations at the 2-loop level are used to analyse perturbativity and stability of the vacuum. We impose experimental and theoretical constraints on the model and find regions in the parameter space consistent with the dark matter relic abundance inferred from the Planck data and bounds on DM-nucleon scattering cross-section from XENON and LUX experiments.
\end{abstract}
\section{Introduction}
The Standard Model of particle physics (SM) is an extremely successful theory, which describes the properties of all known elementary particles, notably also the characteristics of the Higgs boson. Nevertheless, the cosmology requires also the existence of the dark matter (DM), which cannot be built out of the fields included in the SM, therefore this is the major motivation to explore its extensions.

In this paper we study a model of a Higgs portal with a complex scalar field responsible for the mass generation of an abelian vector dark matter. We check, if it can explain the value of the DM relic abundance obtained from the measurements of the Planck satellite and fulfill constraints coming from colliders and DM direct detection experiments.   

We also consider the~issue of the vacuum stability and discuss whether the metastability of the scalar potential, which is a feature of the SM \cite{Buttazzo:2013uya}, can be avoided in the extended framework. In particular, we present the modifications to the issue of stability due to the existence of the vector dark matter. 

\section{Description of the model and theoretical constraints}
\label{sec:vdm}

The model of vector dark matter (VDM) is an extension of the SM with a complex scalar field $S$ that is charged under an extra $U(1)_X$ gauge factor and has nonzero vacuum expectation value (VEV) \cite{Lebedev:2011iq}. Consequently, an extra massive gauge boson appears in the theory, which can be a DM candidate, provided it is stable. This can be ensured by imposing $Z_2$	symmetry on the $U(1)_X$ boson $A_\mu$ with the following transformation rules:
\beq
A^{\mu}_X \rightarrow -A^{\mu}_X \ , \ S\rightarrow S^*,  \ \text{where} \  S=\phi e^{i\sigma}, \ \ {\rm so} \ \ \phi\rightarrow \phi,  \ \  \sigma\rightarrow -\sigma.
\label{symmetry}
\eeq
As a result the kinetic mixing of $U(1)_X$ and the SM hypercharge $U(1)_Y$ is forbidden, whereas the $U(1)_Y$ \& $SU(2)_L$ gauge bosons mix as in the SM. The masses of vectors are given by:
\beq
M_W=gv/2, \ \ \ \ M_Z = \sqrt{g^2+g'^2} v/2 \ \ \ \text{and} \ \ \   \mzp = \gx \vx,
\eeq
where $v$ and $\vx$ are VEVs of $H$ and $S$ respectively and $g_x$ is the $U(1)_X$ gauge coupling.
The scalar potential in this model can be written as
\beq
V= -\mu^2_H|H|^2 +\lambda_H |H|^4 -\mu^2_S|S|^2 +\lambda_S |S|^4 +\kappa |S|^2|H|^2.
\label{potential}
\eeq
It has three positivity conditions which are imposed in further discussions: 
\beq
\lambda_H > 0, \ \ \lambda_S >0, \ \ \kappa > -2 \sqrt{\lambda_H \lambda_S}.
\label{positivity}
\eeq

The mass squared matrix $\mathcal{M}^2$ for the fluctuations $ \left(\phi_H, \phi_S\right)$ of the scalar fields around VEVs can be diagonalized via rotation by a mixing angle $\alpha$
\begin{equation} 
\mathcal{M}^2 = \left( 
\begin{array}{cc}
2 \lambda_H v^2  & \kappa v \vx \\ 
 \kappa v \vx &2 \lambda_S v^2_x 
\end{array} 
\right), 
\left(\begin{array}{c}
h_1\\ 
h_2
\end{array} 
\right)=\left( 
\begin{array}{cc}
\cos \alpha   & -\sin \alpha \\ 
\sin \alpha &\cos \alpha
\end{array} 
\right)
\left(
\begin{array}{c}
\phi_H\\ 
\phi_S
\end{array} 
\right)
\label{massmatrix}
 \end{equation} 

We chose the convention in which $h_1$ is the observed Higgs particle. The~potential has 5 parameters, but using the constraints $\mone=125.7$~GeV and $v=246.22$~GeV, we can eliminate two of them and, including $g_x$, use four independent parameters ($\lambda_H$, $\lambda_S$, $\kappa$, $g_x$) to describe the model.
\section{Stability of the vacuum}
In order to check the stability and perturbativity conditions we adopted the renormalisation group equations (RGE). There is a large quantitative difference in the running of parameters between 1- and 2-loop levels. Therefore, we used more accurate 2-loop approximation. It was obtained with the help of the \SARAH software \cite{Staub:2013tta} and cross-checked using \PYRATE \cite{Lyonnet:2013dna}. 

\begin{figure}[h]\centering
\includegraphics[width=1\textwidth]{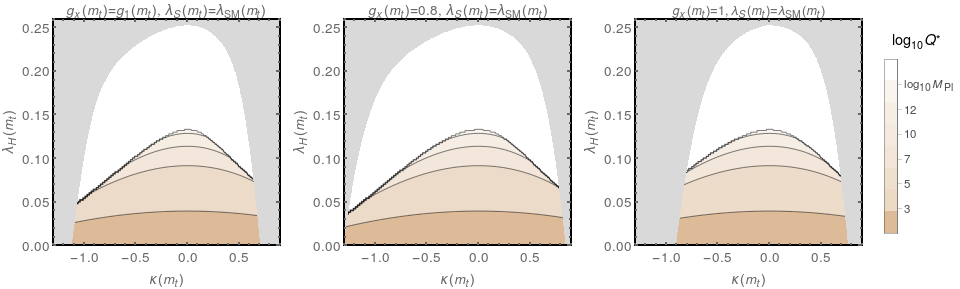}\\
\includegraphics[width=1\textwidth]{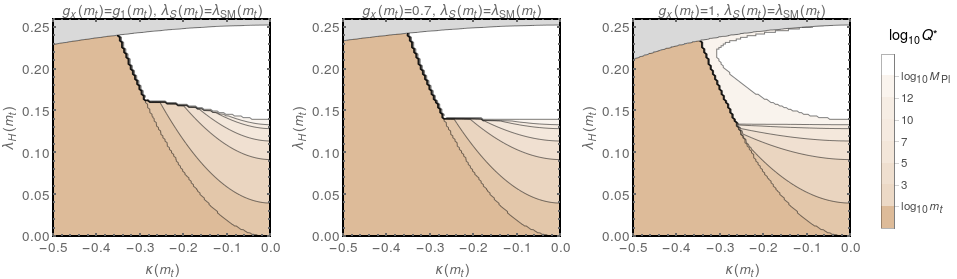}\\
\caption{Contours of constant energy scale $Q^*$ at which the first (upper panel) or the third (lower panel) of the positivity conditions (\ref{positivity}) is violated. In the grey areas quartic couplings become non-perturbative below the Planck scale.
}
\label{posplots}
\end{figure}

The stability of the vacuum was analysed by checking the positivity conditions for quartic couplings (\ref{positivity}), running with	 energy up to the Planck scale. The beta function of $\lambda_S$ was found to be always positive, therefore only two of those conditions are non-trivial (fig. \ref{posplots}). The possibility to alleviate the problem of vacuum stability, in this model, comes mainly from the extra freedom (that is not present in the SM) to choose parameters of the extended scalar potential. If $|\kappa|$ at low energy scale is large enough, then $\lambda_H$ is positive at all scales even if low initial values were chosen. However for $\kappa<0$, the third condition requires 
$\lambda_h(m_T)$ to be larger. The stability regions grows for medium values of $g_x$ and diminishes for large, but only moderately, as this coupling is present exclusively in the beta function of $\lambda_S$ and affects the running of others indirectly.  
\section{Experimental bounds}
Properties of the Higgs within the VDM model agree with those of the SM, however some variation within the limits of experimental and theoretical uncertainties is allowed. Here there is a factor ($\cos\alpha$) that equally suppresses all the couplings of the observed Higgs. It is constrained by the global signal strength $\mu$.
The recent combined analysis of the ATLAS and CMS data \cite{ATLAS} gives a following value of $\mu$ and resulting limit on $\sin\alpha$
\begin{equation}
 \mu=1.09\pm{0.11},\;\;\;\;\;\sin\alpha<0.36, \;\;\;@95\%\;\text{CL}.
\end{equation}
In the mass range $12<\mtwo<90\gev$ stronger bounds come from the  Higgs production process $e^{+}e^{-}\rightarrow Z h_2$ at LEP. Finally measurements of ATLAS restricts the invisible Higgs branching ratio to values $\text{BR}_{\text{inv}}<0.23$ at $95\%$ CL \cite{Aad:2015uga}. The branching ratio $h_1\rightarrow Z'Z',h_2h_2$ dominate the Higgs decays, unless respective couplings $g_x$ or $\kappa$ are tiny. Consequently we exclude regions in the parameter space, where these decays are kinematically allowed.
\begin{figure}[h]\centering
{\includegraphics[width=.49\textwidth,height=0.22\textheight]{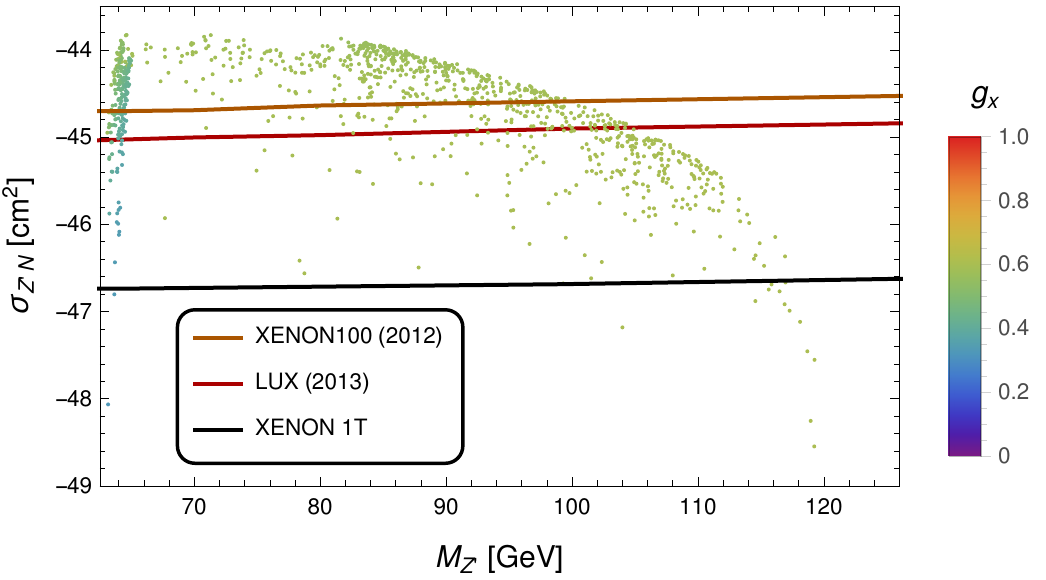}}
{\includegraphics[width=.49\textwidth,height=0.22\textheight]{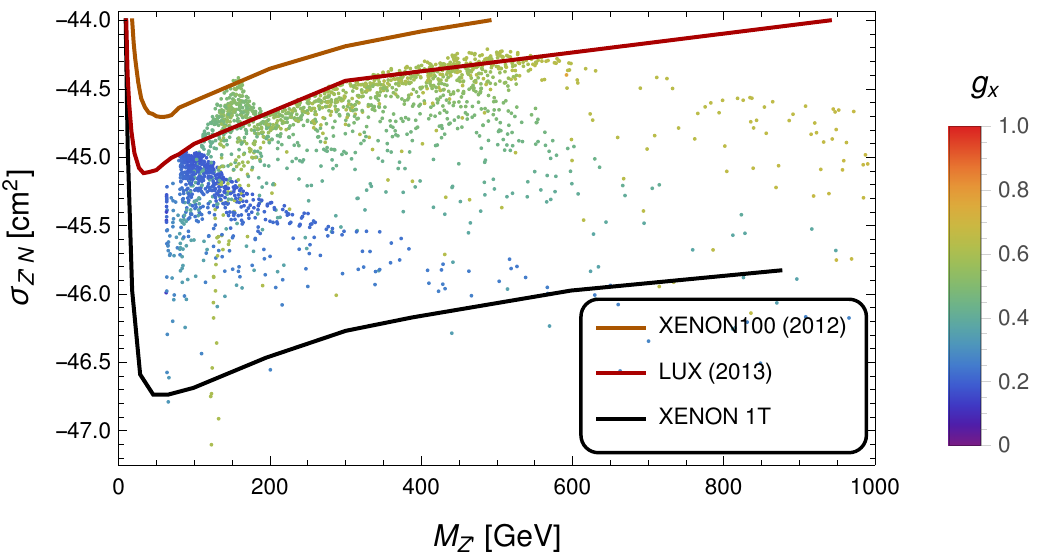}}
\caption{\label{DD-mass}DM - nucleon scattering cross section versus mass of the dark matter for points in the parameter space satisfying all other experimental constraints. 
}
\end{figure}

We assume that the dark matter was thermally produced in the early universe and compute its relic abundance with the use of micrOMEGAs\cite{Belanger:2013oya}. Dark matter - nucleon
scattering is mediated by Higgs particles and their mass-degeneracy leads to destructive interference in the cross section
\vspace{-0.18cm}
\beq
\xdd=\frac{\bar{\mu}^2}{4\pi}\gx^2 g_{hNN}^2 \sin^2 2 \alpha \left({M_{h_1}^{-2}}-M_{h_2}^{-2}\right)^2,
\eeq
where $\bar{\mu}$ is the reduced mass and $g_{hNN}$ is the effective Higgs to nucleon coupling \cite{Cheng2014}. We performed a scan in the parameter space and found points that are within $5\sigma$ limit from the Planck value \cite{Ade:2015xua} of the relic abundance $\Omega_{DM} h^2=0.1199\pm 0.0022$ and satisfy all the collider constraints (fig. \ref{DD-mass}).
\vspace{-0.3cm}
\section{Vector dark matter in multi-component scenario}
It is worth to mention, that it is possible to extend the single VDM scenario discussed so far to a multi-component dark matter model. It can be done by adding an extra dark matter field $\chi$ charged under $U(1)_X$ (and neutral under SM symmetries), which couples minimally to $Z'$. The vector boson remains stable, if the decay into $\chi\chi$ is kinematically forbidden ($M_{Z'}<2M_\chi$), whereas other decays of $Z'$ and~$\chi$ are disallowed by the symmetry~(\ref{symmetry}) and conservation of $\chi$ charge. If~the second DM candidate is a fermion, its Majorana mass can be generated via term $S\chi^TC\chi + H. c.$ provided its charge equals $-1/2$ of the scalar~$S$ charge. The  field $\chi$ communicates with the SM mainly via $S$. Details of the model will be discussed elsewhere.
\section{Summary and conclusions}
The vector dark matter model is a theory, which can explain the observed DM relic abundance. It leads to the attractive prediction of the second Higgs boson and allows to alleviate the problem of the vacuum stability. This model can be further tested by sensitive DM direct detection experiments and at LHC by refined measurements of the Higgs or searches for other scalars. A~two-component extension of the model is possible.
\section{Acknowledgements}
This work has been supported in part by the National Science Centre
(Poland) as  research projects no DEC-2014/15/B/ST2/00108 and DEC-2014/13/B/ST2/03969.
\vspace{-0.5cm	}


\end{document}